\documentclass{ACFD}
\begin{document}
\title{Assessment of Pressure Based Solver in Resolving Complex Shock Wave Phenomenon}
\author{Kuldeep Singh Rawat$^{1*}$, 	Swapnil Ahire$^1$, 	Avijit Chatterjee$^{1}$}
\affiliation{
		$^1$ Department of Aerospace Engineering, Indian Institute of Technology Bombay, Mumbai, India\\
		\vspace{7pt}
%  	     	$^2$ University/Company, Institute/Department/Division, City, Country\\
		\vspace{7pt}
		$^*$ ae.kuldeep@gmail.com
		}
\maketitle
%%%%%%%%%%%%%%%%%%%%%%%%%%%%%%%%%%%%%%%%%%%%%%%%%%
%%%%%%%%%%%%%%%%%%%%%%%%%%%%%%%%%%%%%%%%%%%%%%%%%%
\begin{abstract}
This study presents a critical assessment of a pressure-based solver (PBS) in resolving complex interactions of shocks, turbulent structures etc.. The canonical problem chosen to be resolved in this study is of mode staging in axisymmetric supersonic jet screech. The screech phenomenon exhibits staging behavior characterized by frequency and azimuthal structure changes at specific frequencies. The PBS simulations in the popular ANSYS Fluent software-suite were validated against numerical work and experimental measurements, and results were analyzed. Simulations are performed on supersonic jets which emits dual high frequency screech tones at particular Mach numbers. At lower end of these supersonic Mach numbers, the flow can involve vanishingly weak shock strengths which is routinely captured in experiments and by density based solvers in literature. The limitations of the pressure-based solver in resolving complex shock flow phenomena and predicting mode staging are highlighted at vanishingly weak Mach numbers, emphasizing the need for further investigation given the recent popularity of such solvers for all Mach numbers including in high-speed flow. \\[1em]
\end{abstract}
%%%%%%%%%%%%%%%%%%%%%%%%%%%%%%%%%%%%%%%%%%%%%%%%%%
\section{Introduction}
In the development of computational fluid dynamics (CFD), different approaches have been employed to simulate fluid flows with varying characteristics. The pressure-based approach was initially developed for low-speed incompressible flows, while the density-based approach was used for high-speed compressible flows\cite{fluent2020ansys}. Using preconditioning techniques, density-based solvers can effectively handle a wide range of flows, including low-speed incompressible flows\cite{shen2014implementation}. Pressure based solvers with dual-time stepping, local preconditioning and artificial compressibility methods have proven effective for the simulation of  compressible flows at all Mach numbers\cite{nguyen2023review}. Coupled pressure-based methods have been validated on various CFD benchmark cases for supersonic flow with shocks and now forms a popular choice for high speed compressible flow simulation in common software packages \cite{darwish2014fully, chen2010coupled}. However, a critical assessment of PBS in resolving complex shock wave phenomenon is missing in literature. The majority of users now view such solvers in CFD software packages as a black box, making it particularly crucial to address this aspect.
% This is especially important with such solvers in CFD software package being progressively \textbf{missing word} as a "black box" by majority of users.

%%%%%%%%%%%%%%%% 
\subsection{Popularity of Pressure Based Solver}
ANSYS Fluent and OpenFOAM are two popular CFD software packages in commercial and open source domain respectively that incorporate a range of pressure-velocity coupling algorithms for compressible flow simulations\cite{mangani2016comparing}. Work in reference \cite{kurbatskii2010numerical} focuses on the numerical simulation of axisymmetric supersonic jet screech with pressure based approach using ANSYS Fluent. The work shows that numerical predictions closely match the experimental measurements for the A1 screech mode. However, a discrepancy is reported between the numerical and experimental results for the staging Mach number, particularly for the A2 mode and further investigation is recommended to understand the behavior of the A2 mode \cite{kurbatskii2010numerical}. In the present work, a numerical study was conducted using the popular ANSYS Fluent software package to provide a critical assessment of a generic PBS in resolving complex shock wave flow problem given the recent popularity of such solvers for high-speed flows. The simulations involving supersonic jet screech were validated against experimental and numerical data, and the results were analyzed to determine the validity of using PBS for mode staging behavior which can be considered a canonical problem for complex shock wave related phenomenon.
%%%%%%%%%%%%%%%%
\section{Jet Screech}
Screech tones are produced by imperfectly expanded jets under certain conditions \cite{raman1998advances}. Jet screech is a high-pitched noise at a very specific frequency. The screech sound is characterized by a high-frequency tone that can be as loud as 140 decibels and is often described in literature as a sharp, piercing sound. 

%%%%%%%%%%%%%%%%
\subsection{Mode Staging in Axisymmetric Jet}
The screech phenomenon exhibits staging behavior at certain Mach numbers, characterized by changes in frequency and downstream moving azimuthal structures \cite{edgington2019aeroacoustic}. At lower Mach numbers, axisymmetric toroidal screech modes A1 and A2 are observed. Several theories have been proposed to predict the screech frequency and mode staging behavior. Most involve modifications to Powell's formula including that by Li and Gao \cite{gao2010multi} which is found to provide fairly accurate predictions. However, there is no general theory to explain mode staging behavior and it remains an open problem in supersonic aerodynamics \cite{edgington2022unifying}. In computer simulations, the Discrete Fourier Transform (DFT) is used to compute acoustic frequencies present in the near flow field by analyzing the pressure history recorded at specific points. It is crucial for a solver to accurately identify the two distinct frequency peaks at the A1 and A2 modes to capture the mode staging process which happens at specific jet exit Mach numbers.
%%%%%%%%%%%%%%%%
\section{Numerical Methodology}
Supersonic jet screech was initially analyzed by URANS simulation in ANSYS Fluent using a PBS. Gradients needed were calculated using the Green-Gauss node-based gradient evaluation. For the spatial discretization of pressure, a second order scheme was used, while for the other governing equations, QUICK scheme was used. A second order implicit time marching scheme is used to march the solution in time. An axisymmetric hybrid quadrilateral-triangular computational mesh was used, with a mesh size of 148,137 cells based on grid refinement studies. Among these cells, 80.7\% were quadrilateral cells and 19.3\% were triangular cells, while the domain extended to approximately 2000D where D is reference nozzle's exit diameter. Pressure inlet, pressure outlet, and slip (in-viscid) conditions are assigned to inlet, far-field and walls. Pressure outlet condition imposed at far-field and large domain size along-with coarse mesh in far-field ensures no spurious reflections from the outlet boundary. To refine the grid, a local grid refinement study was conducted, focusing on the near field of the nozzle, spanning 5D radially and 20D axially. This region exhibits significant flow gradients due to boundary layers, shock structures, and their interactions. Solution convergence was deemed attained when time history of acoustic pressure signals monitored at two microphone location on the nozzle lip consistently maintained a stable oscillatory state.
\begin{figure}[h]
        \centering
    	 \includegraphics[width=.75\textwidth]{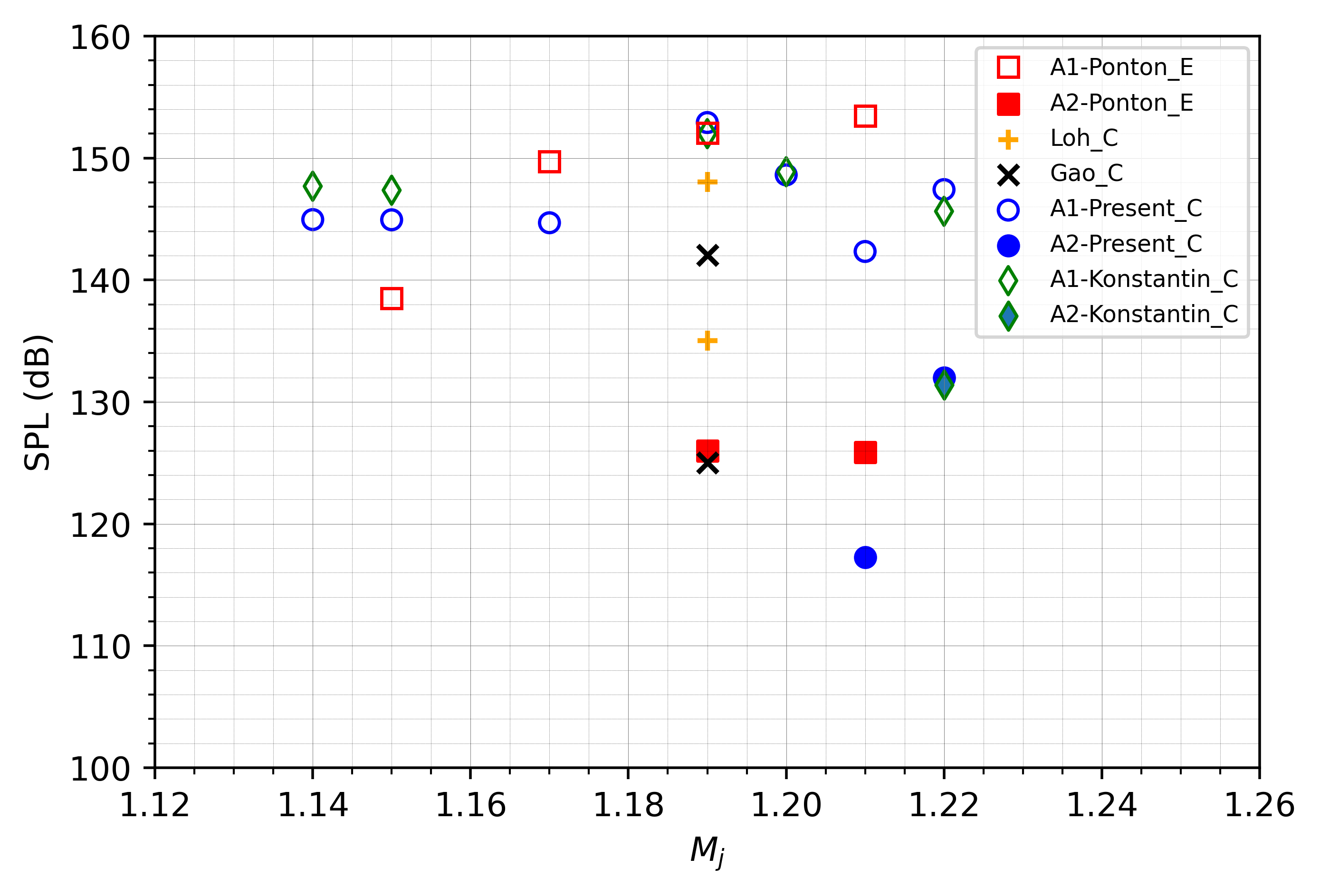}
	 \caption{SPL vs $M_{j}$: Comparison with experiment \cite{ponton1997near} and numerical \cite{kurbatskii2010numerical, gao2010multi, loh2001near} : C is for numerical and E is for experimental data}
        \label{fig:1}
\end{figure}

%%%%%%%%%%%%%%%% 
\section{Results and Validation}
\begin{figure}[h]
        \centering
    	 \includegraphics[width=.60\textwidth]{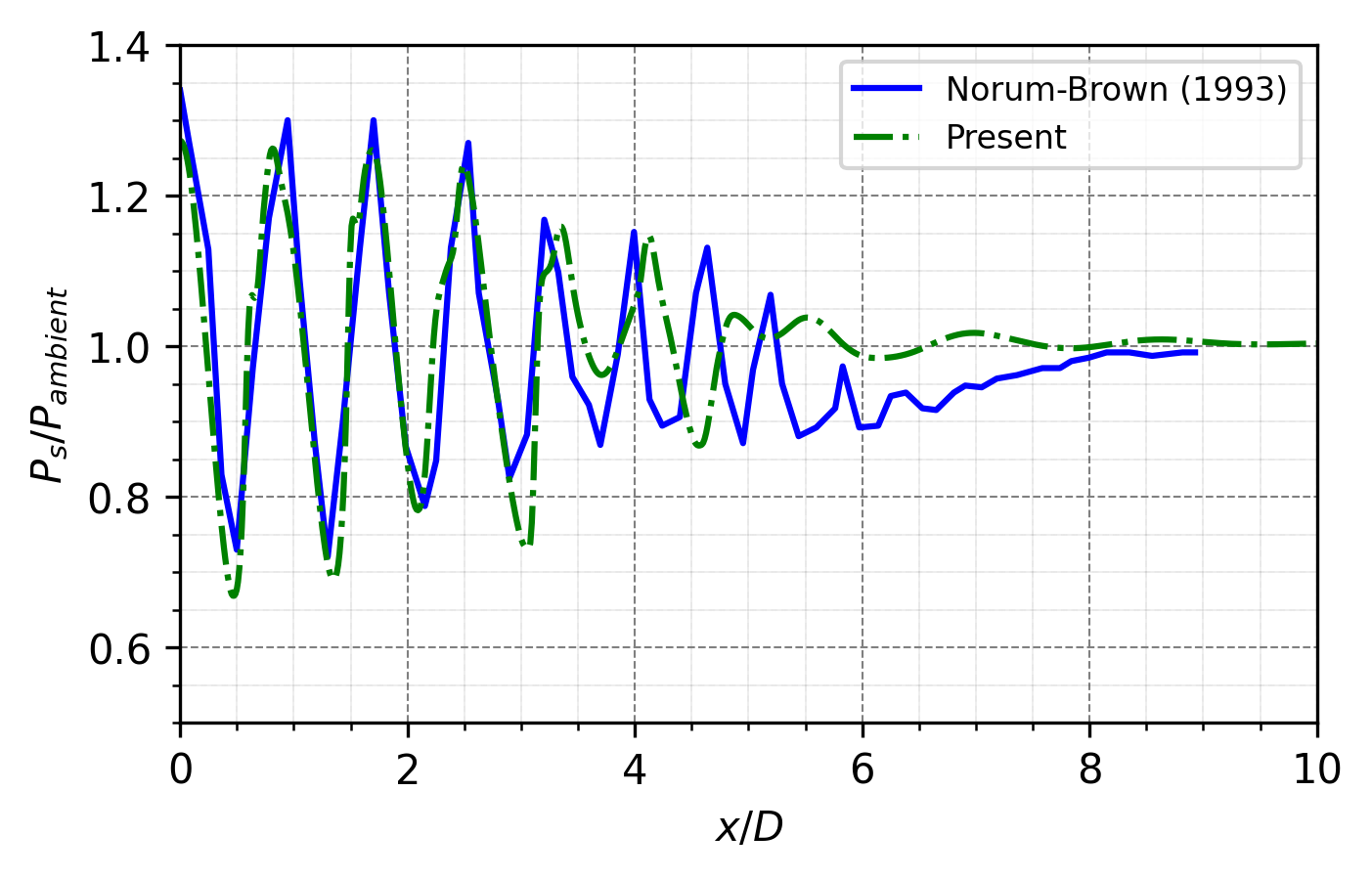}
	 \caption{Comparison of numerical and experimental\cite{norum1993simulated} time-averaged centerline pressure distributions for $M_{j}$ = 1.20}
        \label{fig:2}
\end{figure}

Reference nozzle of 1.0$''$ exit diameter was chosen for this study. Both experimental and computational data is readily accessible for this nozzle configuration. Initial validation results were obtained for fully expanded Mach numbers ($M_j$) of 1.15, 1.19, and 1.20, for reference nozzle's lip thickness of 0.625$''$. These results were compared with numerical data \cite{kurbatskii2010numerical, gao2010multi, loh2001near} and experimental data \cite{ponton1997near}.The results included  sound pressure level (SPL) versus frequency as in Fig. \ref{fig:1} and averaged centerline pressure as shown in Fig. \ref{fig:2}, which were found to be in good agreement with the reference. Screech tones, which are linked to jet instabilities, can impact the shock structure, thereby influencing the patterns of pressure drop variations as seen in Fig. \ref{fig:2}.

\begin{figure}[htbp]
    \centering
    \subfloat[Noise spectra for $M_{j}$ = 1.19]{\includegraphics[width=0.50\textwidth]{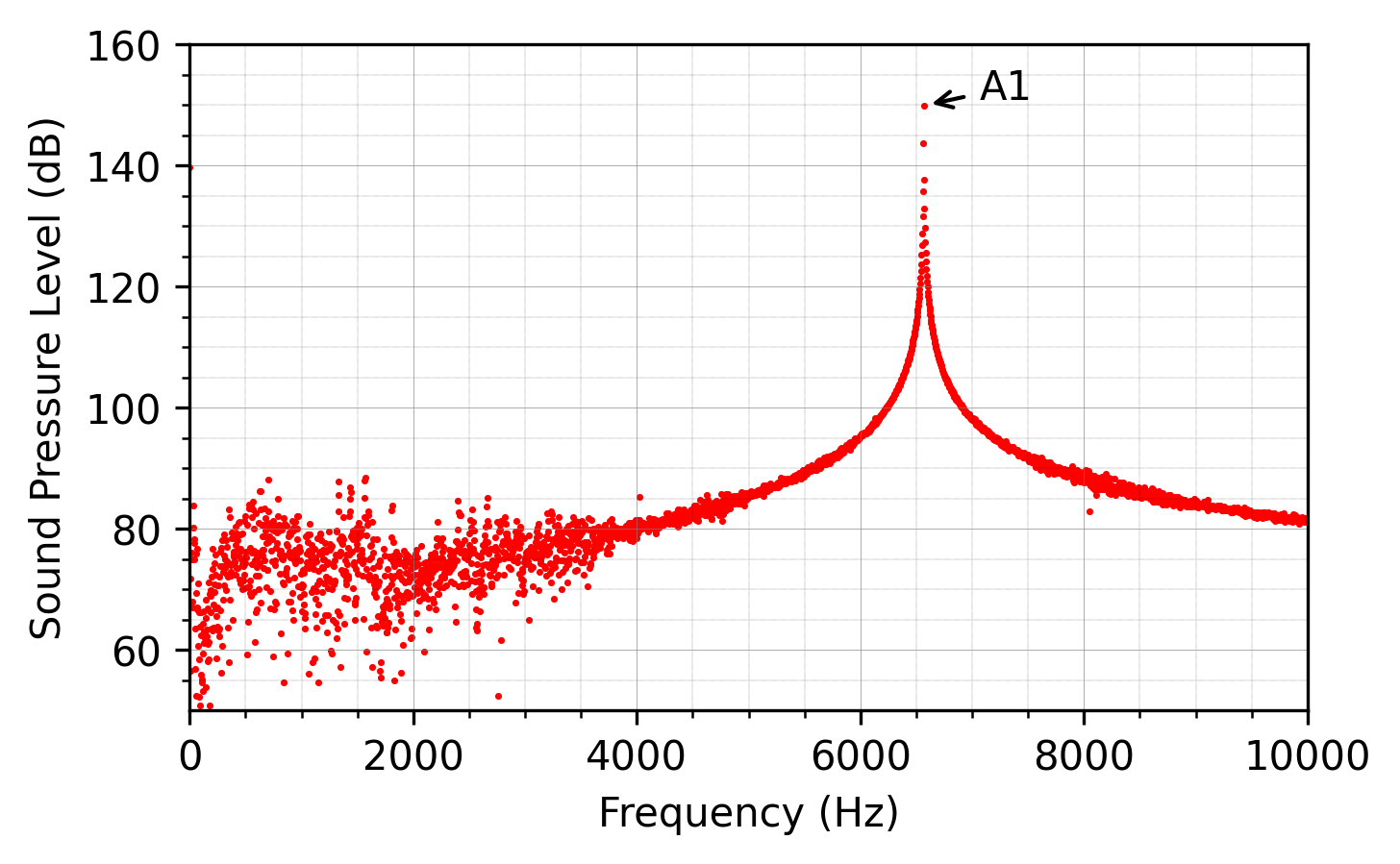}\label{fig:3a}}
    \hfill
    \subfloat[Noise spectra for $M_{j}$ = 1.22]{\includegraphics[width=0.50\textwidth]{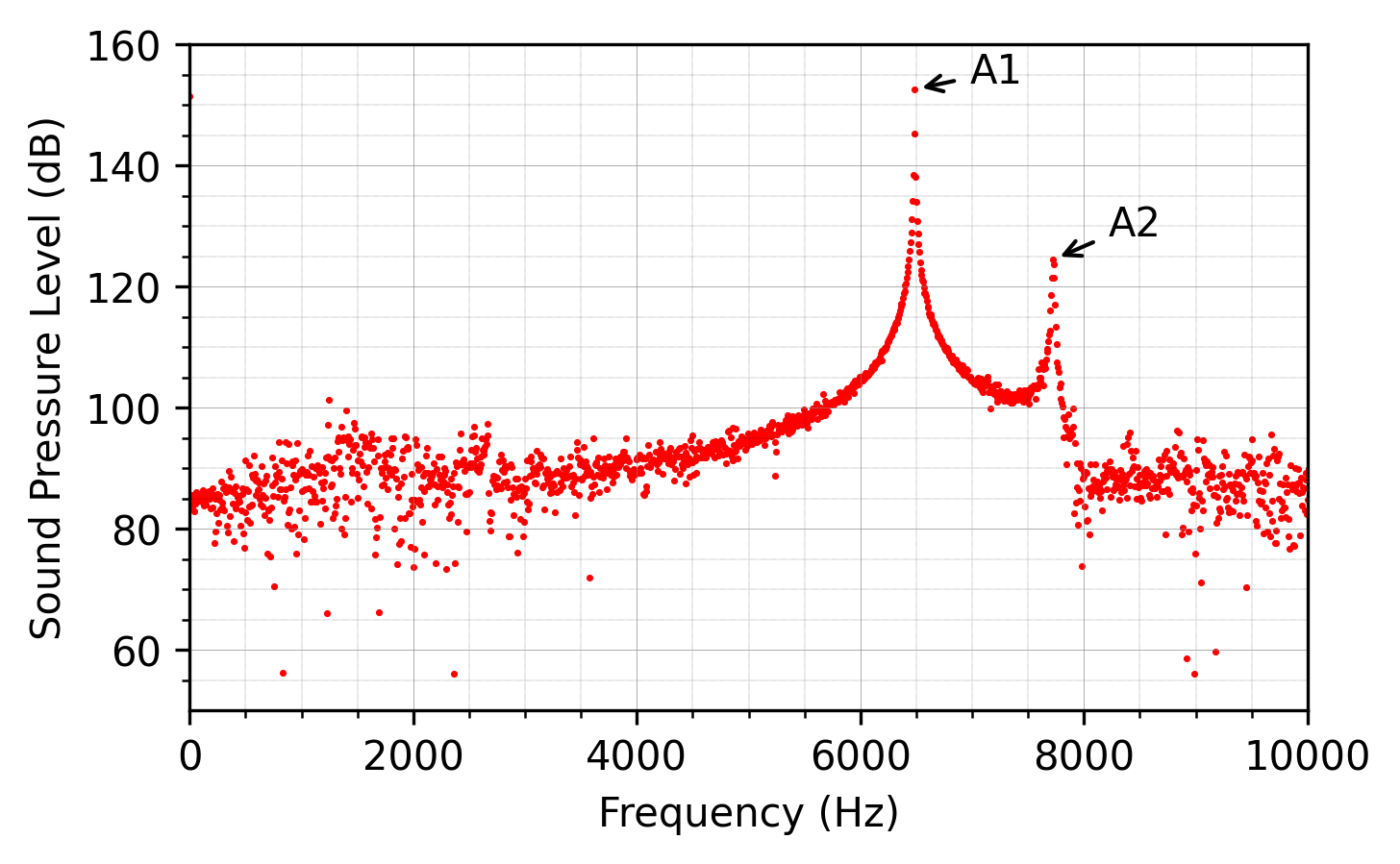}\label{fig:3b}}
    \caption{Spectral analysis of pressure signal}
    \label{fig:3}
\end{figure}

\begin{figure}[htbp]
    \centering
    \subfloat[Numerical time-averaged centerline pressure distributions]{\includegraphics[width=0.50\textwidth]{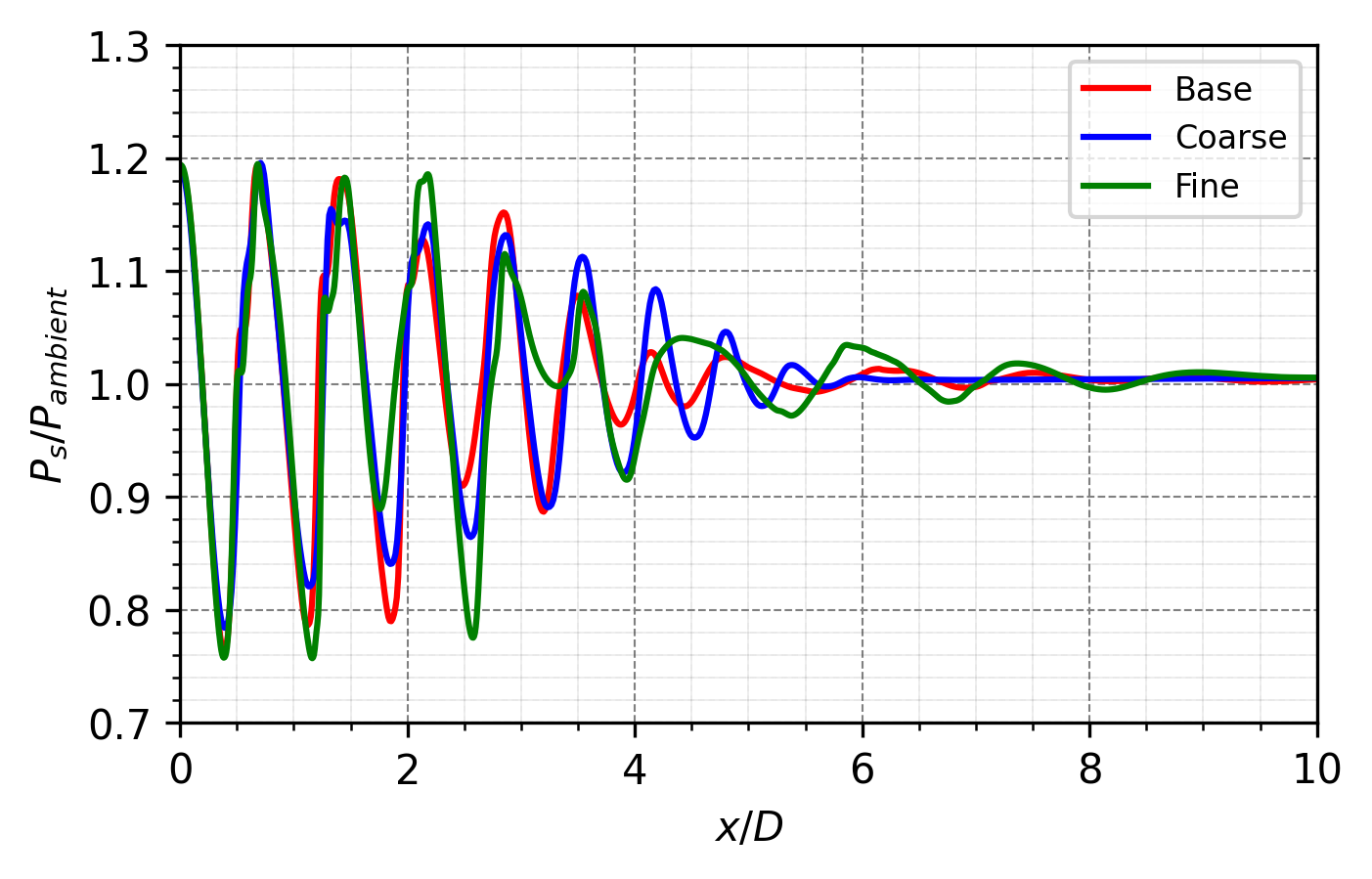}\label{fig:4a}}
    \hfill
    \subfloat[Comparison of SPL and frequency]{\includegraphics[width=0.50\textwidth]{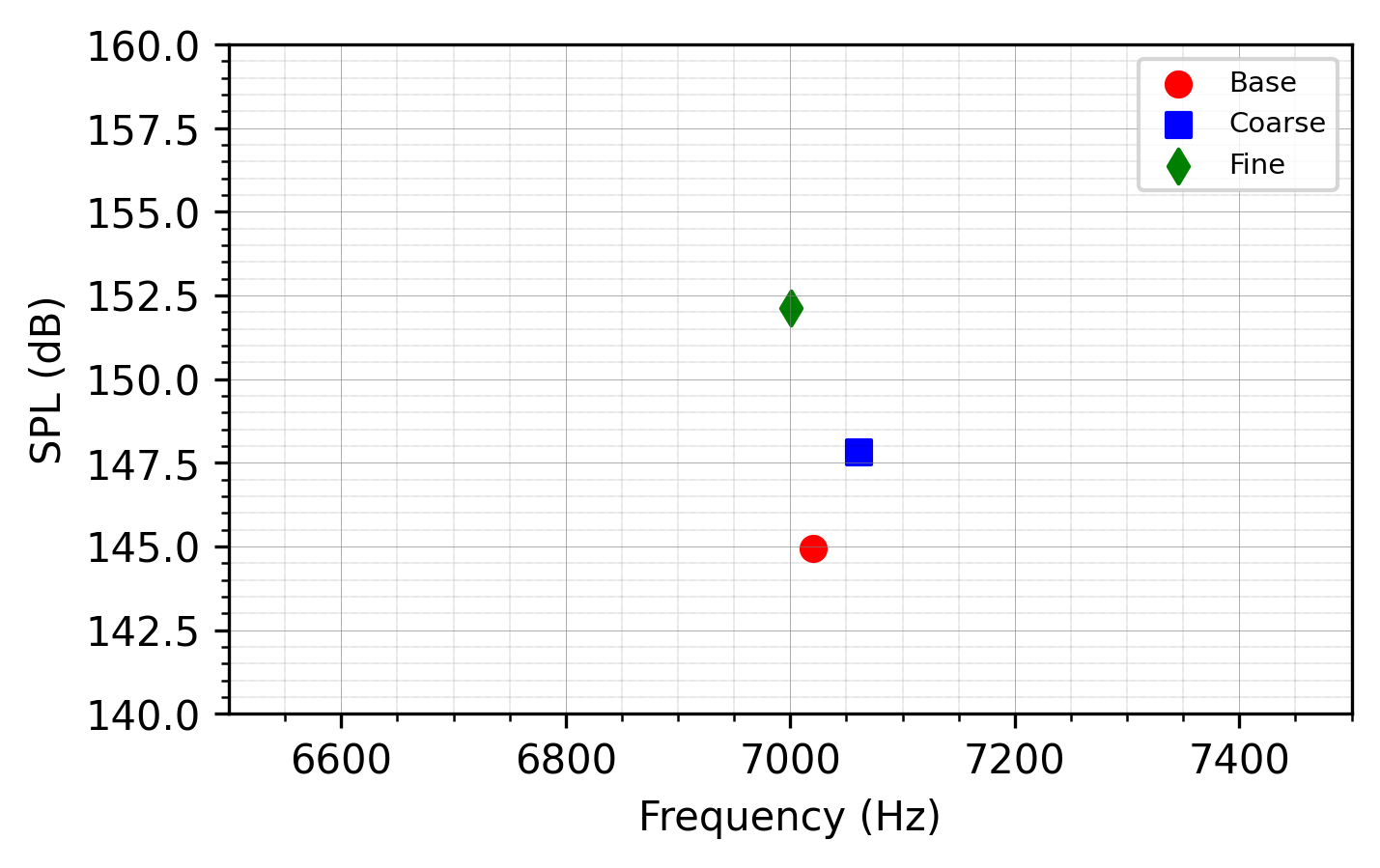}\label{fig:4b}}
    \caption{Grid Independence study for $M_{j}$ = 1.15}
    \label{fig:4}
\end{figure}

Figure \ref{fig:1} compares SPL at one microphone location on the nozzle lip for $M_{j}$ mentioned above. The results obtained for fundamental screech mode A1 are in close agreement with existing reference data. Additional Mach numbers were subsequently simulated and are plotted in Fig. \ref{fig:1}. In the present numerical study, A2 mode is observed at $M_{j}$ = 1.21 and 1.22 as shown in Fig. \ref{fig:1} and Fig. \ref{fig:3b}. However, A2 mode is also expected at $M_{j}$ = 1.19, which is not captured by the PBS as shown in Fig. \ref{fig:3a}. To further validate the approach, a grid independence study was conducted for $M_j$ = 1.15. Figure \ref{fig:4a} shows that solutions from different grid sizes overlap and exhibit negligible differences upto $4^\text{th}$ shock cell which is indicative of adequately resolved spatial scales. Figure \ref{fig:4b} shows SPL and frequency for three different grids are within 5\% and 0.6\% respectively with respect to base grid. The consistent SPL values across grids and the close agreement in frequencies suggest that the simulation effectively captures the flow's temporal dynamics.

\begin{figure}[h]
        \centering
    	 \includegraphics[width=0.65\textwidth]{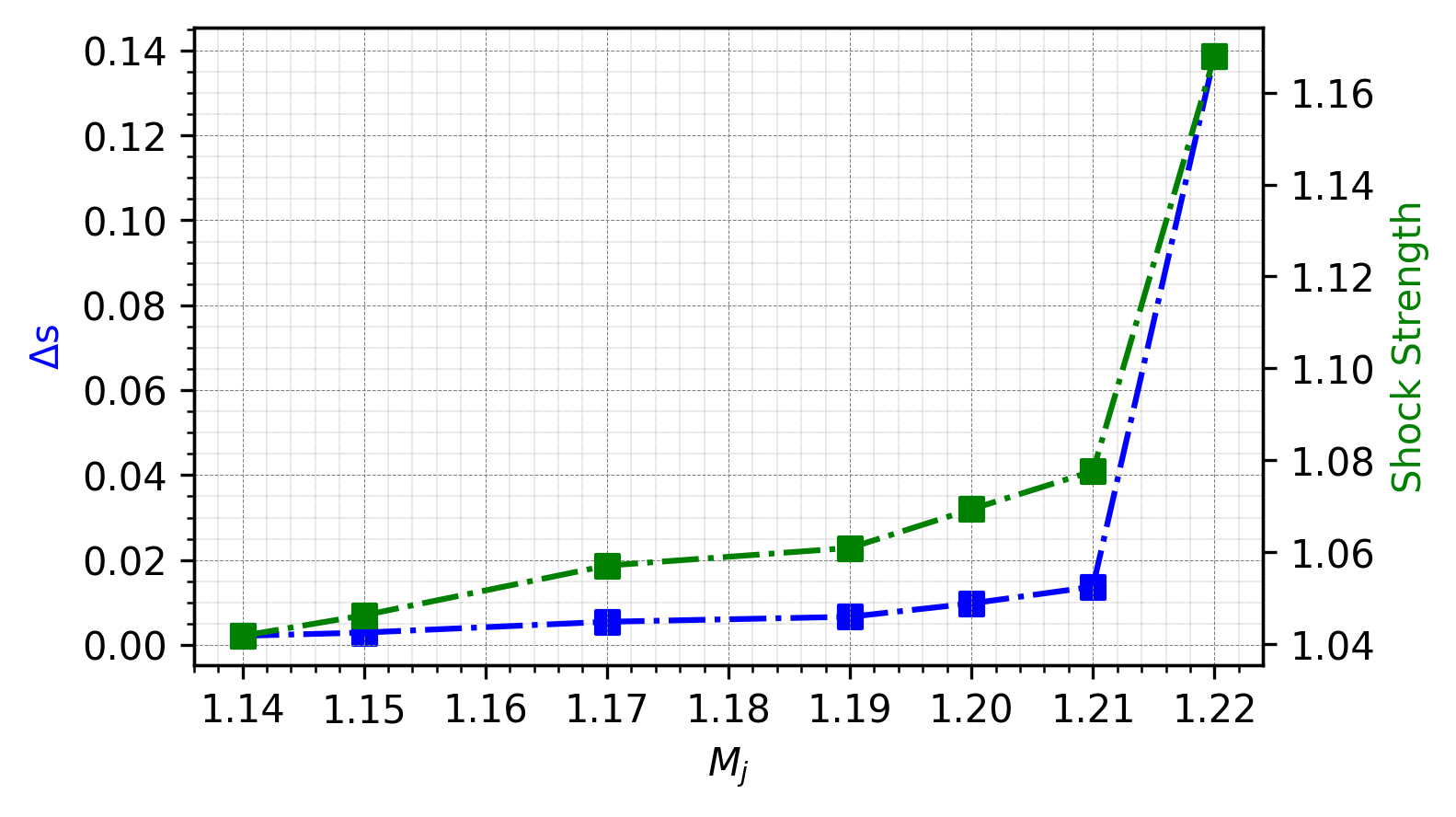}
	 \caption{Variation of change in entropy and shock strength with $M_{j}$}
        \label{fig:5}
\end{figure}

Further assessment of PBS was carried out for $M_{j}$ = 1.19 which exhibits mode staging in literature. Different turbulence models, higher order spatial schemes and a finer time step were tested. Simulation with DES (Detached Eddy Simulation) predicted the jet structure but high frequency screech tones were not present. Simulation with k-$\omega$ model was able to predict two discrete high frequencies in the range of 5kHz to 10 kHz but SPL levels were much lower compared to literature. Convergence issues were also encountered with higher order spatial schemes. This suggests that the current approach using a PBS predicts mode staging at relatively higher $M_{j}$ when shock strength in the screeching jet is higher. Figure \ref{fig:5} compares the shock strength ($P_{2}$/$P_{1}$) and entropy change ($\Delta$s) across the shock wave of strength $M_{j}$ close to unity based on well known 3rd order relation of $\Delta$s with $\Delta$P across weak shock waves \cite{Book1}. The shock strengths in the supersonic screeching jet have an upper bound of $M_{j}$ to account for the over or under expansion. At $M_{j}\sim$  1.19, the almost isentropic nature of the vanishingly weak shock wave requires the underlying solver to be sensitive enough to capture the complex mode staging behavior. PBS seems to fail in this respect compared to density based solvers to reproduce mode staging observed in experiments at lower Mach number of the screeching jets.

%%%%%%%%%%%%%%%% 
\section{Conclusions}
Despite the recent popularity of PBS proved in commercial as well as open source domain for simulating high-speed flow, assessment of such solvers to resolve complex shock wave phenomenon is missing in literature. Mode staging behavior in axisymmetric supersonic jet screech is taken to be the canonical problem representing complex shock wave phenomenon in present study. The analysis indicates an inability of PBS to resolve complex shock wave phenomenon as in mode staging seen in supersonic jet screech at low supersonic Mach numbers which are routinely captured by density based solvers.

%%%%%%%%%%%%%%%% 
%\begin{acknowledgments}
%Put acknowledgments here.
%\end{acknowledgments}
%%%%%%%%%%%%%%%% 
%%%%%%%%%%%%%%%% 
% put bibliography here
\bibliography{ref}
%\printbibliography

\end{document}